\documentclass{article}

\usepackage{arxiv}

\usepackage{graphicx}
\usepackage{url}
\usepackage{hyperref}
\usepackage[numbers,sort&compress]{natbib}
\usepackage{amsmath}
\usepackage{nicefrac}       % compact symbols for 1/2, etc.
\usepackage[utf8]{inputenc}
\usepackage[T1]{fontenc}
\usepackage{appendix}
\usepackage{subcaption} % for sub figures

\newcommand{\aim}{quantify the conditions under which a typical particle filter is able to reliably estimate the `true' state of an underlying pedestrian system through the combination of a modelled state estimate, produced using an agent-based model, and observational data}

\title{Simulating Crowds in Real Time with Agent-Based Modelling and a Particle Filter
\thanks{This project has received funding from the European Research Council (ERC) under the European Union’s Horizon 2020 research and innovation programme (grant agreement No. 757455), through a UK Economic and Social Research Council (ESRC) Future Research Leaders grant [number ES/L009900/1], and through an internship funded by the UK Leeds Institute for Data Analytics (LIDA).}}

\author{
  Nick Malleson\thanks{Corresponding author}\\
  School of Geography\\
  University of Leeds\\
  Leeds, LS2 9JT, UK\\
  \textit{n.s.malleson@leeds.ac.uk}\\
 \And 
 Kevin Minors\\
  Leeds Institute for Data Analytics\\
  University of Leeds\\
  Leeds, LS2 9JT, UK \\
   \And 
 Le-Minh Kieu\\
  Leeds Institute for Data Analytics\\
  University of Leeds\\
  Leeds, LS2 9JT, UK \\
   \And 
 Jonathan A. Ward\\
  School of Mathematics\\
  University of Leeds\\
  Leeds, LS2 9JT, UK \\
   \And 
 Andrew A. West\\
  School of Geography\\
  University of Leeds\\
  Leeds, LS2 9JT, UK \\
   \And 
 Alison Heppenstall\\
  Alan Turing Institute\\
  British Library\\
  London NW1 2DB \\
 } 

\begin{document}
\maketitle              % typeset the header of the contribution
\begin{abstract}
Agent-based modelling is a valuable approach for systems whose behaviour is driven by the interactions between distinct entities. They have shown particular promise as a means of modelling crowds of people in streets, public transport terminals, stadiums, etc. However, the methodology faces a fundamental difficulty: there are no established mechanisms for dynamically incorporating \textit{real-time} data into models. This limits simulations that are inherently dynamic, such as pedestrian movements, to scenario testing of, for example, the potential impacts of new architectural configurations on movements. This paper begins to address this fundamental gap by demonstrating how a particle filter could be used to incorporate real data into an agent-based model of pedestrian movements at run time. The experiments show that it is indeed possible to use a particle filter to perform online (real time) model optimisation.  However, as the number of agents increases, the number of individual particles (and hence the computational complexity) required increases exponentially. By laying the groundwork for the real-time simulation of crowd movements, this paper has implications for the management of complex environments (both nationally and internationally) such as transportation hubs, hospitals, shopping centres, etc.

\end{abstract}


% !TEX root = ParticleFilter.tex
\section{Introduction\label{introduction}}

Agent-based modelling is a form of computer simulation that is well suited to modelling human systems~\citep{bonabeau_agent_2002,  farmer_economy_2009}. In recent years it has emerged as an important tool for decision makers who need to base their decisions on the behaviour of crowds of people~\citep{henein_agentbased_2005}. Such models, that simulate the behaviour of synthetic individual people (`agents'), have been proven to be useful as tools to experiment with strategies for humanitarian assistance \citep{crooks_gis_2013}, emergency evacuations \citep{ren_agentbased_2009, schoenharl_design_2011}, religious festivals~\citep{zainuddin_simulating_2009}, crowd stampedes~\citep{helbing_simulating_2000} etc. Although many agent-based crowd simulations have been developed, there is a fundamental methodological difficulty that modellers face: there are no established mechanisms for incorporating real-time data into simulations \citep{lloyd_exploring_2016, wang_data_2015, ward_dynamic_2016}. Models are typically calibrated once, using historical data, and then projected forward in time to make a prediction independently of any new data that might arise. Although this makes them effective at analysing scenarios to create information that can be useful in the design of crowd management policies, it means that they cannot currently be used to simulate real crowd systems \textit{in real time}. Without knowledge of the current state of a system it is difficult to decide on the most appropriate management plan for emerging situations.

Fortunately, methods do exist to reliably incorporate emerging data into models. \textit{Data assimilation} (DA) is a technique that has been widely used in fields such as meteorology, hydrology and oceanography, and is one of the main reasons that weather forecasts have improved so substantially in recent decades \citep{kalnay_atmospheric_2003}. Broadly, DA refers to a suite of techniques that allow observational data from the real world to be incorporated into models \citep{lewis_dynamic_2006}. This makes it possible to more accurately represent the current state of the system, and therefore reduce the uncertainty in future predictions.

It is important to note the differences between the data assimilation approach used here and that of typical agent-based parameter estimation / calibration. The field of optimisation -- finding suitable estimates for the parameters of algorithms -- has (and continues to be) an extremely well-researched field that agent-based modellers often draw on. For example, agent-based models regularly make use of sampling methods, such as Latin Hypercube sampling \citep{thiele_facilitating_2014} or evolutionary / heuristic optimisation algorithms such as simulated annealing~\citep{pennisi_optimal_2008}, genetic algorithms,~\citep{heppenstall_genetic_2007}, and approximate Bayesian computation~\citep{grazzini_bayesian_2017}. There are also new software tools becoming available to support advanced parameter exploration \citep{ozik_extremescale_2018}. It is, however, worth noting that in most cases agent-based models are not calibrated to quantitative data~\citep{thiele_facilitating_2014}.  In the cases where parameter estimation does take place, it is typically performed as a single calibration step in a waterfall-style development process -- e.g. design, implementation, calibration, validation. Although there are some more recent studies that do attempt to re-calibrate model parameters dynamically during runtime \citep[e.g.][]{oloo_adaptive_2017a, oloo_predicting_2018} there is another, more fundamental, difference to typical parameter optimisation (be it static or dynamic) and the data assimilation approach. 

Even if optimal model parameters have been found, there will usually be a degree of uncertainty in the model. With crowd simulations, for example, it is impossible to know \textit{exactly} how individuals will behave in a given situation -- will someone turn left or right given two competing options? -- nor can individual parameters such as walking speed ever be known exactly. Data assimilation algorithms use `state estimation' to calculate the difference between the model and the `true' state of the underlying system at runtime. They are then able to adjust the current model state in order to constrain a model's continued evolution against the real world~\citep{ward_dynamic_2016}. Although it is possible to re-calibrate models dynamically during runtime (e.g. \citep{oloo_adaptive_2017a}), this would not reduce the natural uncertainty that arises as stochastic models evolve. 

This paper is part of a wider programme of work\footnote{\url{http://dust.leeds.ac.uk/}} whose main aim is to develop data assimilation methods that can be used in agent-based modelling. The software codes that underpin the work discussed here are available in full from the project code repository; see Appendix~\ref{appendix:code}. The work here focuses on one particular system -- that of pedestrian movements -- and one particular method -- the particle filter. A particle filter is a brute force Bayesian state estimation method whose goal is to estimate the `true' state of a system, obtained by combining a model estimate with observational data, using an ensemble of model instances called particles. When observational data become available, the algorithm identifies those model instances (particles) whose state is closest to that of the observational data, and then re-samples particles based on this distance. It is worth noting that once an accurate estimate of the \textit{current} state of the system has been calculated, predictions of future states should be much more reliable -- c.f. the substantial improvements in weather forecasting that have come about as a result of modern data assimilation methods~\citep{kalnay_atmospheric_2003}. Predicting future system states is beyond the scope of this paper however.

The overall aim of the paper is to: \begin{quote}\textit{\aim}.\end{quote} This will be achieved through a number of experiments following an `identical twin' approach \citep{wang_data_2015}. The agent-based model is first executed to produce hypothetical real data -- also known as pseudo-truth \citep{grazzini_bayesian_2017} data -- and these data are assumed to be drawn from the real world. During data assimilation, observations are derived from the pseudo-truth data. This approach has the advantage that the `true' system state can be known precisely, and so the accuracy of the particle filter can be calculated. In reality, the true system state can never be known.

The agent-based model under study is designed to represent a very simple pedestrian system. It has been kept intentionally simple because the aim here is to experiment with the particle filter, not to accurately simulate a pedestrian system. Were the model more complicated it would become more difficult to understand the internal uncertainties, which would in turn make it more difficult to understand how well the particle filter was able to handle these uncertainties. The model is sufficiently complex to allow the emergence of crowding, so its dynamics could not be easily replicated by a simpler mathematical model -- as per \citep{lloyd_exploring_2016} and \citep{ward_dynamic_2016} -- but is otherwise as simple as possible. Crowding occurs because the agents have a variable maximum speed, therefore slower agents hold up faster ones who are behind them. The only uncertainty in the model, which the particle filter is tasked with managing, occurs when a faster agent must make a random choice whether to move round a slower agent to the left or right. Without that uncertain behaviour the model would be deterministic. A more realistic crowd simulation \citep[e.g.][]{helbing_simulating_2000} would exhibit much more complicated behavioural dynamics.

The paper is outlined as follows: Section~\ref{background} reviews the relevant literature; Section~\ref{method} outlines the methods, including a description of the agent based model and particle filter; Section~\ref{experiments} outlines the experiments that are conducted and their results; and Section~\ref{discussion} draws conclusions and outlines opportunities for future work.

% !TEX root = ParticleFilter.tex
\section{Background\label{background}}

\subsection{Agent-Based Crowd Modelling}

One of the core strengths of agent-based modelling is its ability to replicate hetereogeneous individuals and their behaviours.  These individuals can be readily placed within an accurate representation of their physical environment, thus creating a potentially powerful tool for managing complex environments such as urban spaces. Understanding how pedestrians are likely to move around different urban environments is an important policy area.  Recent work using agent-based models have shown value in simulating applications from the design of railway stations \citep{klugl_largescale_2007, chen_multiagentbased_2017}, the organisations of queues \citep{kim_modeling_2013}, the dangers of crowding~\citep{helbing_simulating_2000} and the management of emergency evacuations \citep{vanderwal_simulating_2017}.

%Other examples could draw on:
%\begin{itemize}
%	\item PEDFLOW model (ABM of pedestrian behaviour) \url{https://www.researchgate.net/publication/245908302_DEVELOPING_THE_BEHAVIOURAL_RULES_FOR_AN_AGENT-BASED_MODEL_OF_PEDESTRIAN_MOVEMENT}
%	\item -Santos, G., \& Aguirre, B. E. (2004). A critical review of emergency evacuation simulation models.. 
%	-Templeton, A., Drury, J., \& Philippides, A. (2015). From mindless masses to small groups: Conceptualizing collective behavior in crowd modeling. Review of General Psychology, 19(3), 215.
%	Good review paper on evacuation or more general crowd models, showing the need for modelling social factors! (Not my work)
%	
%	\item Bosse, T., Hoogendoorn, M., Klein, M. C., Treur, J., Van Der Wal, C. N., \& Van Wissen, A. (2013). Modelling collective decision making in groups and crowds: Integrating social contagion and interacting emotions, beliefs and intentions. Autonomous Agents and Multi-Agent Systems, 27(1), 52-84.
%	This is one of my research again. Here we validated our crowd decision model including emotions and spread of emotions, intentions and beliefs, against other crowd models in a specific setting: the Dam square incident, 04-05-2010. Well cited article. 
%	
%	\item Zheng, X., Zhong, T., \& Liu, M. (2009). Modeling crowd evacuation of a building based on seven methodological approaches. Building and Environment, 44(3), 437-445.
%	A good review paper about seven approaches to model crowds evacuating. ABM is one of the approaches (vs fluid dynamics/gas or lattice models, animal experiments, cellular automate, etc). (Not my work). 
%\end{itemize}

However, a drawback with all agent-based crowd simulations -- aside from a handful of exceptions discussed below --  is that they are essentially models of the past. Historical data are used to estimate suitable model parameters and models are evolved forward in time regardless of any new data that might arise. Whilst this approach has value for exploring the dynamics of crowd behaviour, or for general policy or infrastructure planning, it means that models cannot be used to simulate crowd dynamics in \textit{real time}. The drivers of these systems are complex, hence the models are necessarily probabilistic.  This means that a collection of models (or an ensemble) will rapidly diverge from each other and from the real world even under identical starting conditions. This issue has fostered an emerging interest in the means of better associating models with empirical data from the target system \citep[see, e.g.,][]{wang_data_2015, ward_dynamic_2016, long_spatial_2017}.

\subsection{Data-Driven Agent-Based Modelling}

The concept of Data-Driven Agent-Based Modelling (DDABM) emerged from broader work in data-driven application systems~\citep{darema_dynamic_2004} that aims to enhance or refine a model at runtime using new data. One of the most well developed models in this domain is the `WIPER' system~\citep{schoenharl_design_2011}.  This uses streaming data from mobile telephone towers to detect crisis events and model pedestrian behaviour.  When an event is detected, an ensemble of agent-based models are instantiated from streaming data and then validated in order to estimate which ensemble model most closely captured the particular crisis scenario \citep{schoenharl_evaluation_2008}. Although innovative in its use of streaming data, the approach is otherwise consistent with traditional methods for model validation based on historical data \citep{oloo_adaptive_2017a}. Similar attempts have been made to model solar panel adoption~\citep{zhang_datadriven_2015}, rail travel~\citep{othman_datadriven_2015}, crime \citep{lloyd_exploring_2016}, bird flocking \citep{oloo_predicting_2018} and aggregate pedestrian behaviour \citep{ward_dynamic_2016}, but whilst promising, these models contain their own limitations. For example, they either assume that agent behaviours can be proxied by simple regression models \citep{zhang_datadriven_2015} (which will make it impossible to use more advanced behavioural frameworks to encapsulate the more interesting features of agent behaviour), are calibrated manually \citep{othman_datadriven_2015} (which is infeasible in most cases), optimse model parameters dynamically but not the underlying model state (which might have diverged substantially from reality) \citep{oloo_predicting_2018}, or are simple enough to be approximated by an aggregate mathematical model \citep{lloyd_exploring_2016, ward_dynamic_2016}, neglecting the importance of using agent-based modelling in the first place.

There are two studies of direct relevance. The first is that of \citep{lueck_who_2019}, who developed an agent-based model of an evacuation coupled with a particle filter to conduct real-time data assimilation. This paper presents a novel mathematical approach that can map observations from a simple to a complex domain. In the authors' example, simulated data from hypothetical population counters are mapped to the complex agent-based model, which represents the heterogeneous locations of the individual agents. Although it is beyond the scope of the study here, the proposed mapping method will be useful for future work that experiments with variations in the complexity of the observations that are drawn from the real world (e.g. examining how well the particle filter can perform when presented with aggregate population counts rather than individual agent locations).

The second study is that of \citep{wang_data_2015} who investigated the viability of simulating individual movements in an indoor environment using streams of real-time sensor data to perform dynamic state estimation. As with this paper, the authors used an `identical twin' experimental framework; the agent-based model is used to create pseudo-realistic data rather than using those from real-world sensors. Next, an ensemble of models were developed to represent the target system with a particle filter used to constrain the models to the hypothetical reality. A new particle resampling method (`component set resampling') was also proposed that is shown to mitigate the particle deprivation problem (see Section~\ref{da_pf} for more details). The research presented within this paper builds on \citep{wang_data_2015} by: (i) attempting to apply data assimilation to a system that exhibits emergence; and (ii) performing more rigorous experiments to assess the conditions under which a particle filter is appropriate for assimilating data into agent-based crowd models.

\subsection{Data Assimilation and the Particle Filter\label{da_pf}}

 `Data assimilation' refers to a suite of mathematical approaches that are capable of using up-to-date data to adjust the state of a running model, allowing it to more accurately represent the current state of the target system. They have been successfully used in fields such as meteorology, where in recent years 7-day weather forecasts have become more accurate than 5-day forecasts were in the 1990s~\citep{bauer_quiet_2015}; partly due to improvements in data assimilation techniques~\citep{kalnay_atmospheric_2003}. The need for data assimilation was initially born out of data scarcity; numerical weather prediction models typically have two orders of magnitude more degrees of freedom than they do observation data. Initially the problem was addressed using interpolation~\citep[e.g.][]{panofsky_objective_1949, charney_dynamic_1951}, but this soon proved insufficient~\citep{kalnay_atmospheric_2003}. The eventual solution was to use a combination of \textit{observational data} and the \textit{predictions of short-range forecasts} (i.e. a model) to create the full initial conditions (the `\textit{first guess}') for a model of the current system state that could then  make forecasts. In effect, the basic premise is that by combining a detailed but uncertain model of the system with sparse but less uncertain data, ``all the available information'' can be used to estimate the true state of the target system~\citep{talagrand_use_1991}. 

A particle filter is only one of many different methods that have been developed to perform data assimilation. Others include the Successive Corrections Method, Optimal Interpolation, 3D-Var, 4D-Var, and various variants of Kalman Filtering \citep{carrassi_data_2018}, but it is beyond the scope of the paper to review all of these methods in detail. The particle filter method is chosen here because they are non-parametric methods and are better suited to performing data assimilation in systems that have non-linear and non-Gaussian behaviour~\citep{long_spatial_2017}, such as agent-based models. In fact, agent-based models are typically formulated as computer programs rather than in the mathematical forms required of many data assimilation methods, such as the Kalman filter and its variants \citep{wang_data_2015}.

% As  \citep[][p 37]{wang_data_2015} argue:\begin{quote}``A unique feature of the agent-based simulation model is that the model is specified by behaviours or rules and lacks the analytic structures (e.g., those in partial differential equation models) from which functional forms of probability distributions can be derived. This makes it difficult to apply conventional state estimation techniques such as Kalman filter and its variances.''\end{quote}

The particle filter is a brute force Bayesian state estimation method. The goal is to estimate a posterior distribution (i.e. the probability that the system is in a particular state conditioned on the observed data) using an ensemble of model instances, called particles. Each particle has an associated weight, normalised to sum to one, that are drawn from a prior distribution. In the data assimilation step (discussed shortly) each particle is confronted with observations from the (pseudo) real system and weights are adjusted depending on the likelihood that a particle could have produced the observations. Unlike most other data assimilation methods, a particle filter does not actually update the internal states of its particles. Instead, the worst performing particles -- those that are least likely to have generated the observations -- are removed from subsequent iterations, whereas better performing particles are duplicated. This has the advantage that, when performing data assimilation on an agent-based model, it is not necessary to derive a means of updating unstructured variables. For example, it is not clear how data assimilation approaches that have been designed for mathematical models that consist of purely numerical values will update models that contain agents with categorical  variables.

A particle filter (PF) $P_t$ at time $t$ with $N$ particles is the set 
\begin{equation}
P_t = \left\{ (x_t^i, w_t^i) : i \in \{1,\dots,N\} \right\},
\end{equation}
where $x_t^i$ is the state of the $i$-th particle with associated weight $w_t^i$, which are subject to the condition $\sum_{i=1}^N w_t^i = 1$ for all $t$. The general method of the particle filter is to use $P_t$ and new information in the form of observations to determine $P_{t+1}$. 

There are many different PF methods. The standard form is the sequential importance sampling (SIS) PF which selects the \emph{weights} using importance sampling \citep{bergman_recursive_1999, doucet_sequential_2000}. The particles are sampled from an importance density \citep{uosaki_nonlinear_2003}. One pitfall of the SIS PF is particle degeneracy. This occurs when the weights of all the particles tend to zero except for one particle which has a weight very close to one. This results in the population of particles being a very poor estimate for the posterior distribution.  One method to prevent particle degeneracy is to resample the \emph{particles}, duplicating particles with a high weight and discarding particles with a low weight. The probability of a particle being resampled is proportional to its weight; known as the sequential importance resampling (SIR) PF. The particles can be resampled in a variety of different ways, including multinomial, stratified, residual, systematic, and component set resampling \citep{liu_sequential_1998, douc_comparison_2005, wang_data_2015}. Although resampling helps to increase the spread of particle weights, it is often not sufficient \citep{carrassi_data_2018}. In a similar manner to particle degeneration in the SIS PF, \textit{particle collapse} can occur in the SIR PF. This occurs when only one particle is resampled so every particle has the same state. One of the main drawbacks with PF methods, as many studies have found \citep[e.g.][]{snyder_obstacles_2008, carrassi_data_2018}, is that the number of particles required to prevent particle degeneracy or collapse grows exponentially with the dimensionality of the model. This is an ongoing problem that will be revisited throughout the paper. It is worth nothing that there are many other PFs including the auxiliary SIR PF and the regularised PF \citep{arulampalam_tutorial_2002}, the merging PF \citep{nakano_merging_2007}, and the resample-move PF \citep{gilks_following_2001}. In Section ~\ref{particle_filter}, we will consider a SIR PF with systematic resampling because it ranks higher in resampling quality and computational simplicity compared to other resampling methods \citep{hol_resampling_2006, douc_comparison_2005}.

% !TEX root = ParticleFilter.tex
\section{Method\label{method}}

\subsection{The Agent-Based Model: StationSim}

\begin{figure}[ht]
\centering
\includegraphics[width=0.5\textwidth]{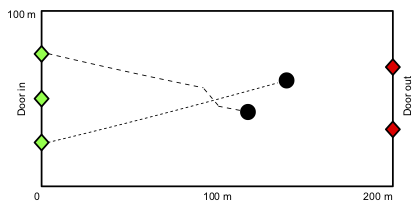}
\caption{StationSim environment with 3 entrance and 2 exit doors}.\label{fig:StationSim}
\end{figure}

\textit{StationSim} is a simple agent-based model that has been designed to very loosely represent the behaviour of a crowd of people moving from an entrance on one side of a rectangular environment to an exit on the other side. This is analogous to a train arriving at a train station and passengers moving across the concourse to leave. A number of agents, $N$, which varies in the later experiments, are created when the model starts. They are able to enter the environment (leave their train) at a uniform rate through one of three entrances. They move across the `concourse' and then leave by one of the two exits. The entrances and exits have a set size, such that only a limited number of agents can pass through them in any given iteration. Once all agents have entered the environment and passed through the concourse then the simulation ends. The model environment is illustrated in Figure \ref{fig:StationSim}, with the trajectories of two interacting agents for illustration. 
%The model is also outlined in full as per the ODD protocol~\citep{grimm_odd_2010} in Appendix~\ref{odd}.

The model has deliberately been designed to be extremely simple and does not attempt to match the behavioural realism offered by more developed crowd models \citep{chen_multiagentbased_2017, helbing_simulating_2000, klugl_largescale_2007, vanderwal_simulating_2017}. The reason for this simplicity is so that: (1) the model can execute relatively quickly; (2) the probabilistic elements in the model are limited (we know precisely from where probabilistic behaviour arises); (3) the model can be described fully using a relatively simple state vector, as discussed in Section~\ref{state_vector}. Importantly, the model is able to capture the emergence of \textit{crowding}. This results because each agent has a different maximum speed that they can travel at. Therefore, when a fast agent approaches a slower one, they attempt to get past by making a random binary choice to move left or right around them. Depending on the agents in the vicinity, this behaviour can start to lead to the formation of crowds. To illustrate this, Figure~\ref{fig:crowding} shows the paths of the agents (\ref{fig:crowding-trails}) and the total agent density (\ref{fig:crowding-density}) during an example simulation.   The degree and location of crowding depends on the random allocation of maximum speeds to agents and their random of direction taken to avoid slower agents; these cannot be estimated \textit{a priori}. Unlike in previous work where the models did not necessarily meet the common criteria that define agent-based models \citep[e.g.]{lloyd_exploring_2016, ward_dynamic_2016} this model respects three of the most important characteristics: 

\begin{figure}
    \centering
    \begin{subfigure}[b]{0.48\textwidth}
        \centering
        \includegraphics[width=.95\linewidth]{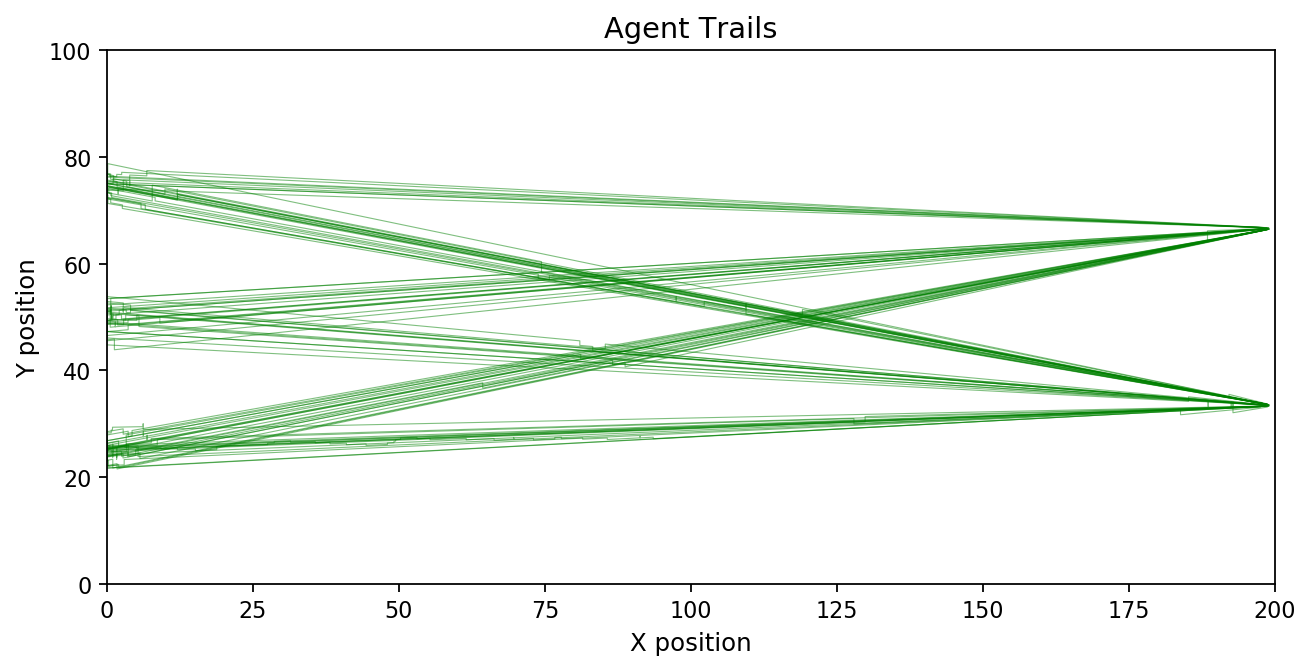}
        \caption{Individual trails showing the paths taken by agents}
        \label{fig:crowding-trails}
    \end{subfigure}
    \label{fig:crowding}
    \begin{subfigure}[b]{0.48\textwidth}
        \centering
        \includegraphics[width=.95\linewidth]{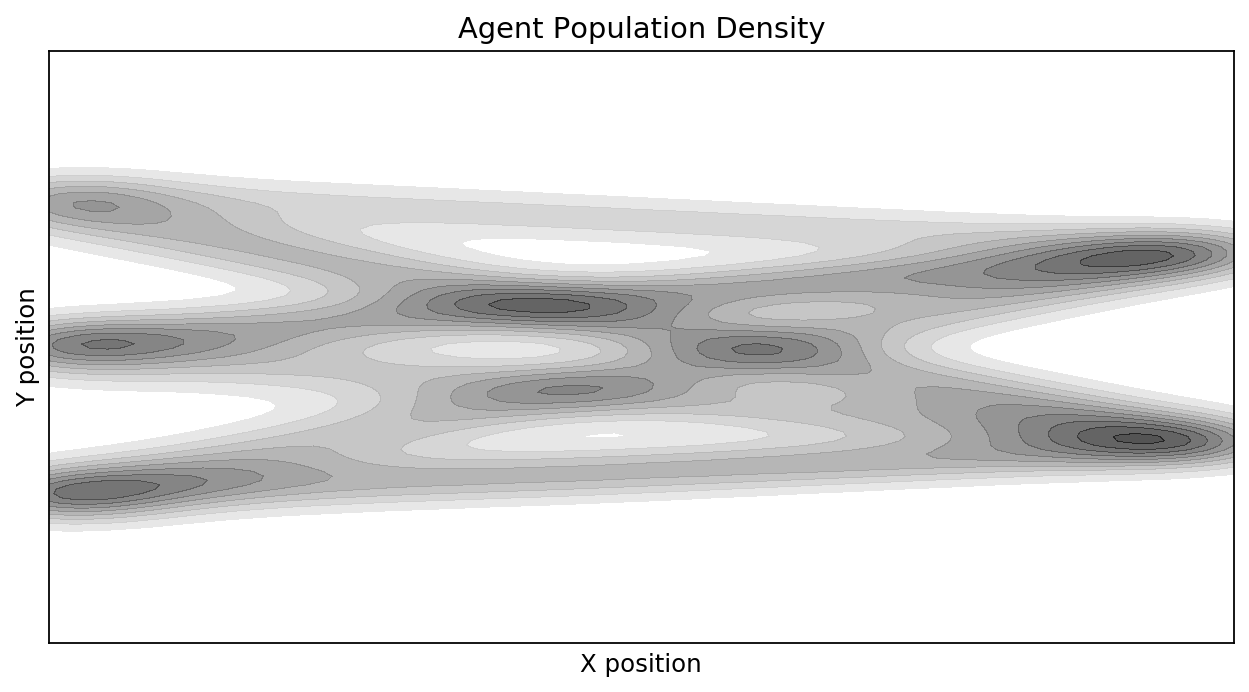}
        \caption{The total crowd density over the simulation run.}
        \label{fig:crowding-density}
    \end{subfigure}
    \caption{An example of crowding in the StationSim model}
    \label{fig:crowding}
\end{figure}

\begin{itemize}
	\item individual heterogeneity -- agents have different maximum travel speeds; 
	\item agent interactions -- agents are not allowed to occupy the same space and try to move around slower agents who are blocking their path; 
	\item emergence -- crowding is an emergent property of the system that arises as a result of the choice of exit that each agent is heading to and their maximum speed.
\end{itemize}

The model code is relatively short and easy to understand. It is written in Python, and is available in its entirety at in the project repository \citep{stationsimgit}.
% \footnote{\url{XXXX Link to repo}}.

\subsection{Data Assimilation - Introduction and Definitions}

DA methods are built on the following assumptions: 

\begin{enumerate}
	\item Although they have low uncertainty, observational data are often spatio-temporally sparse. Therefore there are typically insufficient amounts of data to to describe the system in sufficient detail and a data-driven approach would not work.
	\item Models are not sparse; they can represent the target system in great detail and hence fill in the spatio-temporal gaps in observational data by propagating data from observed to unobserved areas \citep{carrassi_data_2018}. For example, some parts of a building might be more heavily observed than others, so a model that assimilated data from the observed areas might be able to estimate the state of the unobserved areas. However, if the underlying systems are complex, a model will rapidly diverge from the real system in the absence of \textit{up to date} data \citep{ward_dynamic_2016}.
	\item The combination a model and up-to-date observational data allow ``all the available information'' to be used to determine the state of the system as accurately as possible \citep{talagrand_use_1991}. 
\end{enumerate}

DA algorithms work by running a model forward in time up to the point that some new observational data become available. This is typically called the \textit{predict} step. At this point, the algorithm has an estimate of the current system state and its uncertainty (the prior). The next step, \textit{update},  involves using the new observations, and their uncertainties, to update the current state estimate to create a posterior estimate of the state. As the posterior has combined the best guess of the state from the model \textit{and} the best guess of the state from the observations, it should be a closer estimate of the true system state than that which could be estimated from the observations or the model in isolation.

\subsection{The Particle Filter\label{particle_filter}}

\begin{figure}[ht]
\centering
\includegraphics[width=0.8\textwidth]{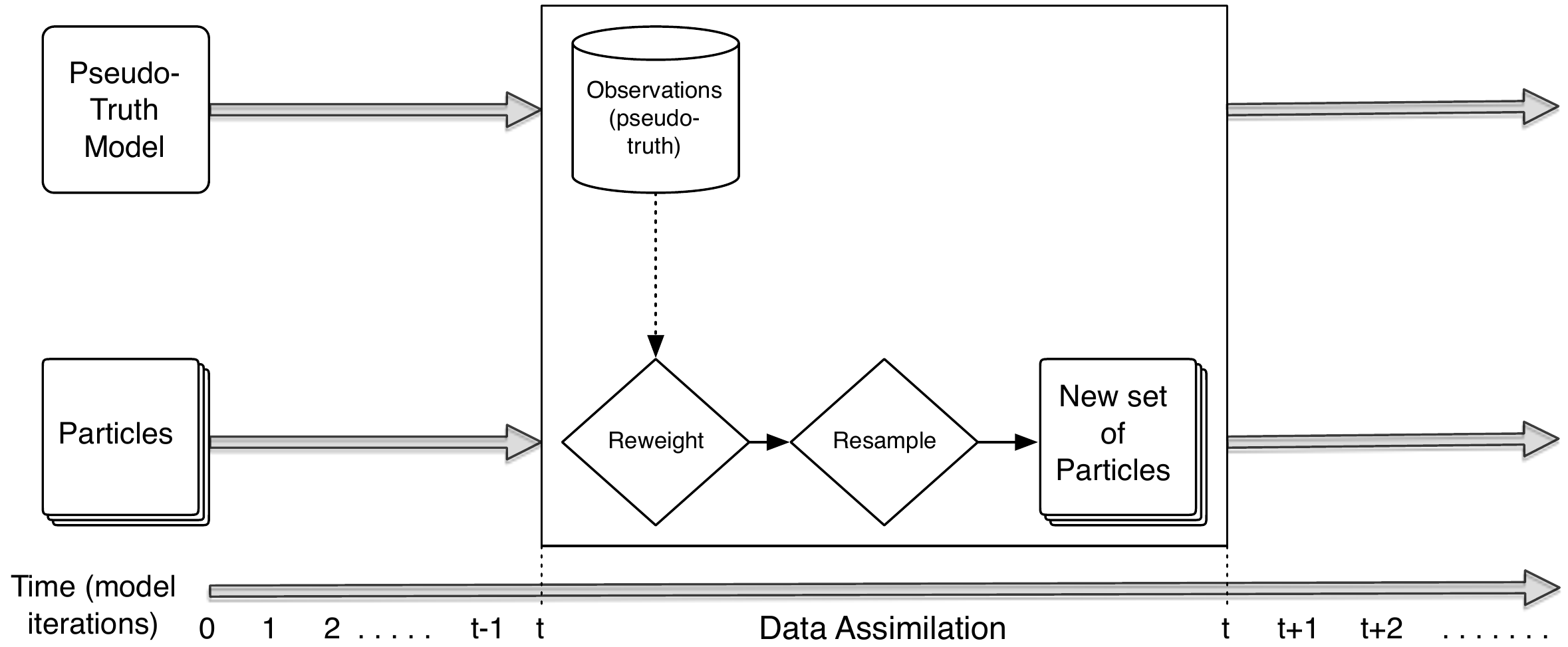}
\caption{Flowchart of data assimilation process using a particle filter.\label{fig:PF_flowchart}}
\end{figure}

There are many different ways to perform data assimilation, as discussed in Section \ref{da_pf}. Here, a potentially appropriate solution to the data assimilation problem for agent-based models is the particle filter -- also known as a Bayesian bootstrap filter or a sequential Monte Carlo method -- which represents the posterior state using a collection of model samples, called particles \citep{gordon_novel_1993,carpenter_improved_1999,wang_data_2015, carrassi_data_2018}. Figure~\ref{fig:PF_flowchart} illustrates the process of running a particle filter. Note that the `pseudo-truth model' is a single instance of the agent-based model that is used as a proxy for the real system as per the identical twin experimental framework that we have adopted.

The data assimilation `window' determines how often new observations are assimilated into the particle filter. The size of the window is an important factor -- larger windows result in the particles deviating further from the real system state -- but here we fix the window at 100 iterations. The simulation terminates when all agents have left the system.

\subsubsection{The State Vector and Transition Function\label{state_vector}} 

Here, the \textit{state vector}, at a time $t$, contains all the information that a transition function needs to iterate the model forward by one step, including all of the agent ($i = \{ 0, 1, \dots, N \} $) parameters ($\overrightarrow{p_i}$) and variables ($\overrightarrow{v_i}$) as well as global model parameters $\overrightarrow{P}$:
\begin{equation}
  S_t  = \left[ \begin{array}{cccccccc}
\overrightarrow{p_0} & \overrightarrow{v_0} & \overrightarrow{p_1} &  \overrightarrow{v_1} &  \dots &  \overrightarrow{p_N} &  \overrightarrow{v_N} & \overrightarrow{P} 
\end{array} \right]
\end{equation} 

A similar structure, the \textit{observation vector}, contains all of the observations made from the `real world' (in this case the pseudo-truth model) at a time $t$. Here, the particle filter is only used to estimate the state of the models variables ($\overrightarrow{v_i}$), not any of the parameters ($\overrightarrow{p_i}$ and $\overrightarrow{P}$) (although it is worth noting that parameter estimation is technically feasible and will be experimented with in future work). Also, the current speed of an agent can be calculated from its current location and the locations of the agents surrounding it, so in effect the observation vector only needs to include the positions of the agents with the addition of some Gaussian noise, $\epsilon$:
\begin{equation}
  O_t  = \left[ \begin{array}{ccccccc}
x_0 & y_0 & x_1 & y_1 & \dots & x_n & y_n 
\end{array} \right]
\end{equation} 

 Therefore in the experiments conducted here, all model parameters are fixed. Hence a further vector is required to map the observations to the state vector that the particles can actually manipulate. We define the partial state vector $S'$ to match the shape of $O$, i.e.:
\begin{equation}
  S'_t  = \left[ \begin{array}{ccccccc}
x_0 & y_0 & x_1 & y_1 & \dots & x_n & y_n 
\end{array} \right]
\end{equation} 

This has the effect of `pairing' agents in the particles to those in the pseudo-truth data, in a similar approach to that taken by \citep{wang_data_2015}. It is worth noting that, because the particle filter will not be tasked with parameter estimation, then the data assimilation is somewhat simpler than it would be in a real application. For example, one of the agent parameters is used to store the location of the exit out of the environment that the agent is moving towards. As this parameter is set \textit{a priori} for each agent, then the particle filter does not need to estimate where the agents are ultimately going, only where they currently are.

\subsubsection{Observations from the pseudo-truth data}

In a real application, the particle filter would be assimilating data in real time from sensors of the real world. This is not the case here so instead, we take `observations' from the pseudo-truth data (which are, as it happens, generated by the StationSim model). Each particle evolves in time according to the StationSim dynamics and receives new observations at regular intervals. Measurement error (i.e. noise, $\epsilon$) is added to the observations (in the real world, sensors observations will be noisy). Here, observations take the form of the $(x,y)$ locations of all agents. This is analogous to tracking all individuals in a crowd and providing snapshots of their locations at discrete points in time to the particle filter. This `synthetic observation' is probably more detailed than reality, so future work will vary the amount of detail provided to the algorithm. It will firstly reduce the number of agents who are observed (e.g. tracking only \textit{some} people) and then provide only aggregate population counts (which is analogous to using a camera or other sensor to count the number of people at a certain point). 

\subsubsection{Particle Weights}

Each particle in the particle filter has a weight associated with it that quantifies how similar a particle is to an observation. The weights are calculated at the end of each data assimilation window (i.e. when observations become available). At the start of the following window the particles then evolve independently from each other \citep{fearnhead_particle_2018}.  The weights are, in effect, the average distance between agents in that particle and the corresponding agents in StationSim (recall that there is a one-to-one mapping between agents in the particles and agents in the truth model). Formally, let $x^n(i,t)$ be the location of the $i$-th agent at time $t$ in the $n$-th particle for $n \in \{1,\dots,N\}$ and let $x(i,t)$ be the location of the $i$-th agent in StationSim for $i \in \{1,\dots,I\}$. The error of the $n$-th particle $\epsilon^n(t)$ at time $t$ is then given by
\begin{equation}
\epsilon^n(t) = \frac{1}{I} \sum_{i=1}^I |x(i,t) - x^n(i,t)|,
\end{equation}
and the particle filter error $\nu(t)$ at time $t$ is given by
\begin{equation}
\label{eqn:particle_error}
\begin{split}
\nu(t) =& \frac{1}{N} \sum_{n=1}^{N} \epsilon^n(t), \\
=& \frac{1}{NI} \sum_{n=1}^{N}\sum_{i=1}^I |x(i,t) - x^n(i,t)|.
\end{split}
\end{equation}

It is worth noting that, because the agent locations are the only data stored in the partial state vector, the particle error is equivalent to the Euclidean distance ($l_2$-norm) between the particle partial state vector $S'_{t}$ and the observation vector $O_t$,
\begin{equation}
 \nu(t) = || S'_{t} - O_t  ||_2.
\end{equation} 

Particles with relatively large error are likely to be removed during the sampling procedure (discussed in the following section), whereas those with low error are likely to be duplicated. In addition for their use in resampling, the population of particle weights can be used to gain insight into how well the particle filter is able to represent the `true' system state overall. 

\subsubsection{Sampling Procedure\label{particle_sampling}}

Here, a bootstrap filter is implemented which uses systematic resampling \citep{doucet_introduction_2001, douc_comparison_2005, wang_data_2015, long_spatial_2017, carrassi_data_2018}. This begins by taking a random sample $U$ from the uniform distribution on the interval $[0,1/N]$ and then selecting $N$ points $U^i$ for $i \in \{1,\dots,N\}$ on the interval $[0,1]$ such that
\begin{equation}
U^i = (i-1)/N + U.
\end{equation}
Let the particles currently have locations $x_i$. Using the inversion method, we calculate the cumulative sum of the normalised particle weights $w_i$ and define the inverse function $D$ of this cumulative sum, that is:
\begin{equation}
D(u) = i \text{ for } u \in \left(\sum_{j=1}^{i-1}w_j,\sum_{j=1}^{i}w_j\right].
\end{equation}
Finally, the new locations of the resampled particles are given by $x_{D(U^i)}$.

As discussed in Section~\ref{background}, a well-studied issue that particle filters face is that of particle deprivation \citep{snyder_obstacles_2008}, which refers to the problem of particles converging to a single point such that all particles, but one, vanish \citep{kong_sequential_1994}.  
% This vastly reduces the size of the state space covered by the population of particles and will make it difficult or impossible to find particles with low error in later windows. Common approaches to resolve the deprivation problem include increasing the number of particles or trying to reduce the dimensionality of the state space. Bespoke approaches also exist; for example \citep{wang_data_2015} develop a technique called `component set resampling' that samples \textit{parts} of particles (i.e. those parts that are working well) rather than whole particles in their entirety.  
Here, the problem is addressed in two ways. Firstly by simply using large numbers of particles relative to the size of the state space and, secondly, by diversifying the particles \citep{vadakkepat_improved_2006} -- also known as roughening, jittering, and diffusing \citep{li_fight_2014, shephard_learning_2009, pantrigo_combining_2005}. In each iteration of the model we add Gaussian white noise to the particles' state vector to increase their variance, which increases particle diversity. This encourages a greater variety of particles to be resampled and therefore makes the algorithm more likely to represent the state of the underlying model. This method is a special case of the resample-move method presented in \citep{gilks_following_2001}. The amount of noise to add is a key hyper-parameter -- too little and it has no effect, too much and the state of the particles moves far away from the true state -- as discussed in the following section. 

% !TEX root = ParticleFilter.tex
\section{Experimental Results\label{experiments}}

\subsection{Outline of the Experiments}

Recall that the aim of the paper is to \aim. By examining the error (i.e. the difference between the particle filter estimate of the system state and the pseudo-truth data) under different conditions, the paper will present some preliminary estimates of the potential for the use of particle filtering in real-time crowd simulation. In particular, the paper will estimate the minimum number of particles ($N_p$) that are needed to model a system that has a given number of agents ($N_a$). Although the dynamics of the model used here are simpler than those of more advanced crowd models, and hence the particle filter is tasked with an easier problem than it would have to solve in a real-world scenario, these experiments provide valuable insight into the general \textit{potential} of the method for crowd simulation.

It is well known that one of the main drawbacks with a particle filter is that the number of particles required can explode as the complexity of the system increases~\citep[e.g.][]{snyder_obstacles_2008, carrassi_data_2018}. In this system, the complexity of the model is determined by the number of agents. As outlined in Section~\ref{method}, randomness is introduced through agent interactions.  Therefore the fewer the number of agents, the smaller the chance of interactions occurring and the lower the complexity. As the number of agents increases,we find that the number of interactions increases exponentially (Figure~\ref{fig:collision_count}, discussed in the following section, will illustrate this). The paper will also experiment with the amount of particle noise that needs to be included to reduce the problem of particle deprivation ($\sigma_p$), but this is not a focus of the experiments.

%Figure~\ref{fig:PF_flowchart} outlined the modelling process. The particle filter runs in parallel to a single instance of the StationSim model, which is used to create pseudo-truth data (observations) representing the the locations of each agent at a given iteration. In each data assimilation `window', the truth model and all particles are stepped forward a number of iterations. At the end of the window the particle filter receives pseudo-true observations and uses these to calculate the particle weights (i.e. the error associated with each particle).  

To quantify the `success' of each parameter configuration -- i.e. the number of particles $N_p$ and amount of particle noise $\sigma_p$ that allow the particle filter to reliably represent a system of $N_a$ agents -- an estimate of the overall error associated with a particular particle filter configuration is required. There are a number of different ways that this error could be calculated. Here, we calculate the mean weight of each particle after resampling in every data assimilation window (the weights vary during an experiment because in early windows there are few agents and hence very little stochasticity so particle errors are low). Then we take the mean of these individual particle errors. In addition, because the results from an individual experiment can vary slightly, each particle configuration is executed a number of times, $M$, and the median error across experiments is calculated.

Formally if $\nu(t,j)$ is the error of particle $j$ in a data assimilation window $t$, and the total number of windows is $T$, then the total error of that particle filter configuration across $M$  experiments is: 
\begin{equation}
  E_{N_p,\sigma_p,N_a} =  {\rm median}_M \left( \sum_{j=1}^{N_p}  \frac{\sum_{t=1}^{T}\nu(t)}{T} \right)
\end{equation}
where $median_M()$ calculates the median over $M$ experiments.

In summary, the Table~\ref{tab:experiment_parameters} outlines the parameters that are used in the experiments. There are a number of other model and particle filter parameters that can vary, but ultimately these do not influence the results outlined here and are not experimented with. The parameters, and code, are available in full from the project repository \cite{stationsimgit}. 
% \begin{quote}\url{XXXX github url}\end{quote}

\begin{table*}[ht] \caption{Main parameters used in the experiments}
	\begin{center}
		\begin{tabular}{l l l } 
		    \hline Parameter & Symbol & Value / Range \\
			\hline
			Number of agents & $N_a$ & $[ 2, 40 ]$ \\
			Number of particles & $N_p$ & $[ 1, 10000 ]$ \\
			Particle noise & $\sigma_p $ & $[0.25, 0.5]$ \\
			Measurement noise & $\sigma_m$ &  $1.0$ \\
			Number of experiments (repetitions) & M & 20 \\
			Model iterations in each data assimilation window & - & 100 \\ 
			\hline
	\end{tabular} \end{center} \label{tab:experiment_parameters}
\end{table*}%

\subsection{Results}

\subsubsection{Overall Error}

\begin{figure}[ht]
	\centering
	\includegraphics[width=0.7\textwidth]{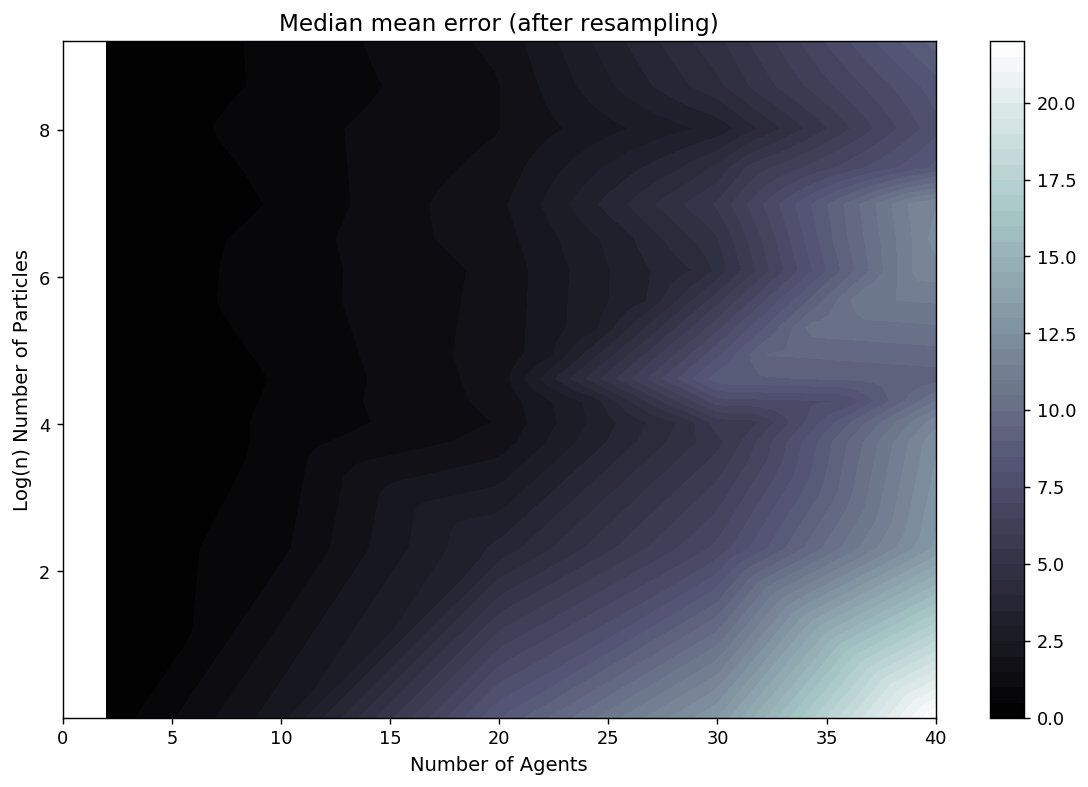}
	\caption{Median of the mean errors after resampling} \label{fig:median_abs_error}
\end{figure}

Figure~\ref{fig:median_abs_error} plots the median of the mean error over all experiments to show how it varies with differing numbers of agents and particles. Due to the computational difficulty in running experiments with large numbers of particles and the need for an exponential increase in the number of particles with the number of agents (discussed in detail below), there are fewer experiments with large numbers of particles and hence the the experiments are not distributed evenly across the agents/particles space. Thus the error is presented in the form of an interpolated heat map. Broadly there is a reduction in error from the bottom-right corner (few particles, many agents) to the top left corner (many particles, few agents). The results illustrate that, as expected, there is a larger error with increasing numbers of agents but this can be mediated with larger numbers of particles. Note the logarithmic scale used on the vertical axis; an exponentially greater number of particles is required for each additional agent included in the simulation. 

\begin{figure}[ht]
	\centering
	\includegraphics[width=0.7\textwidth]{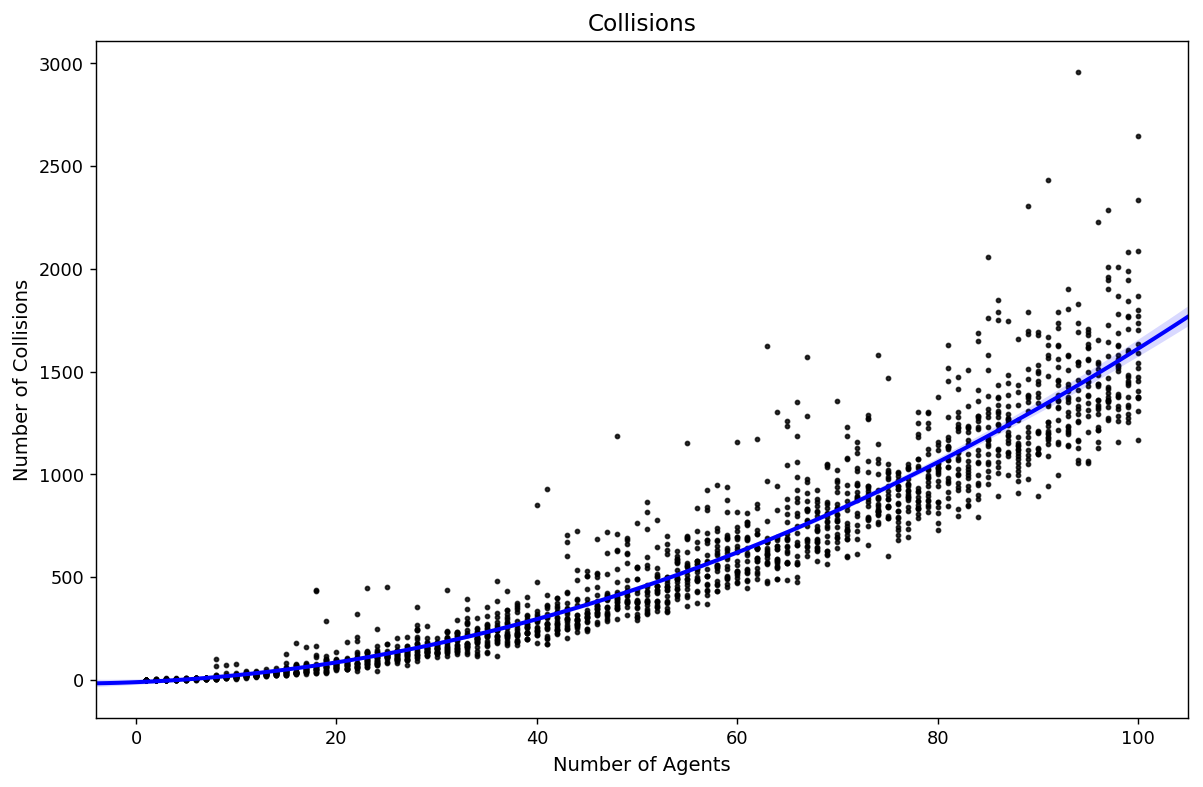}
	\caption{The total number of collisions that occur given the number of agents. The blue line represents a polynomial regression model of order 2 with 99\% confidence intervals.} \label{fig:collision_count}
\end{figure}

There are two reasons for this exponential increase in the number of particles required. Firstly, as the number of agents increases so does the dimensionality of the state space. Also, and perhaps more importantly, with additional agents the chances of collisions, and hence stochastic behaviour, increases exponentially. This is illustrated by Figure~\ref{fig:collision_count} which presents the total number of collisions that occur across a number of simulations with a given number of agents.

\begin{figure}[ht]
	\centering
	\includegraphics[width=0.45\textwidth]{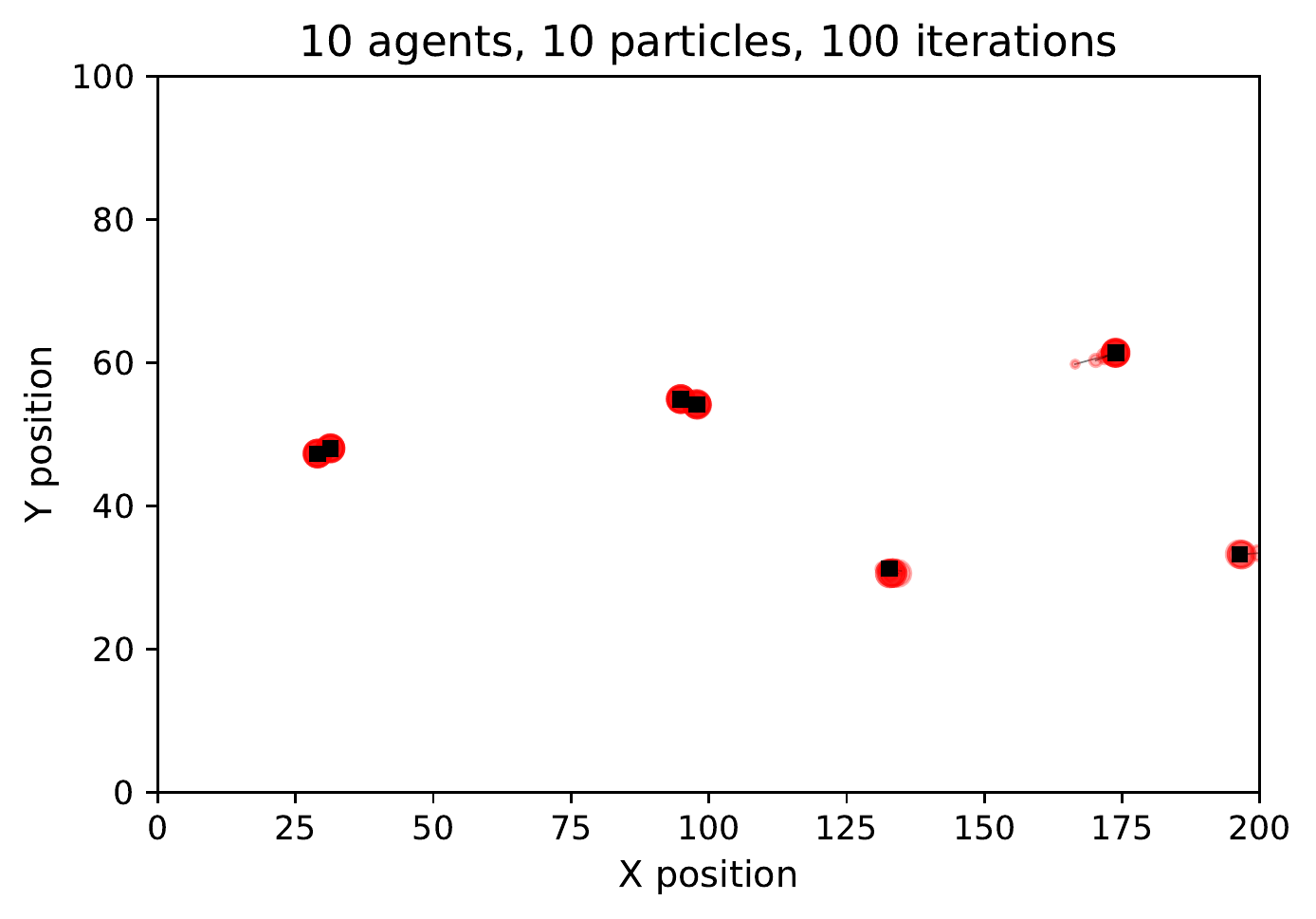}
	\includegraphics[width=0.45\textwidth]{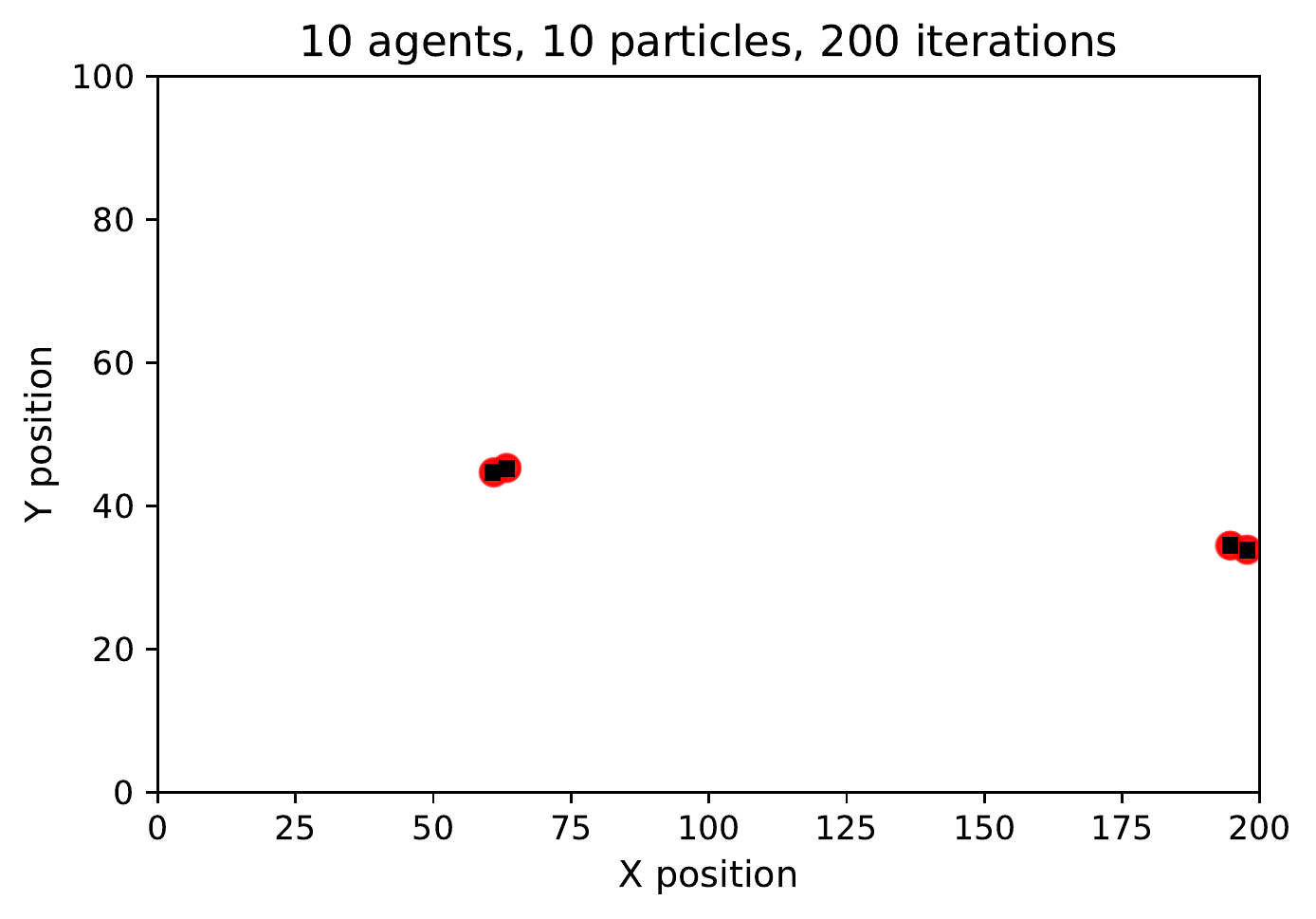}
	\caption{State of a particle filter with 10 agents and 10 particles after 100 and 200 iterations (after resampling at the end of the first and second data assimilation windows). Black squares illustrate the (pseudo) real locations of the agents and red circles represent the locations of those agents as predicted by individual particles (in this case all particles predict the agents' locations accurately and hence the red circles overlap).} \label{fig:ani-10agents-10particles}
\end{figure}

On its own, the overall particle filter error (Figure~\ref{fig:median_abs_error}) reveals little information about which configurations would be judged `sufficiently reliable' to be used in practice. Therefore it is illuminating to visualise some of the results of individual particle filter runs to see how the estimates of the individual agents' locations vary, and what might be considered an `acceptable' estimate of the true state. Figure~\ref{fig:ani-10agents-10particles} illustrates the state of a particle filter with 10 particles and 10 agents at the end of its first data assimilation window (after resampling). With only ten agents in the system there are few, if any, collisions and hence very little stochasticity; all particles are able to represent the locations of the agents accurately.

\begin{figure}[ht]
	\centering
	\includegraphics[width=0.45\textwidth]{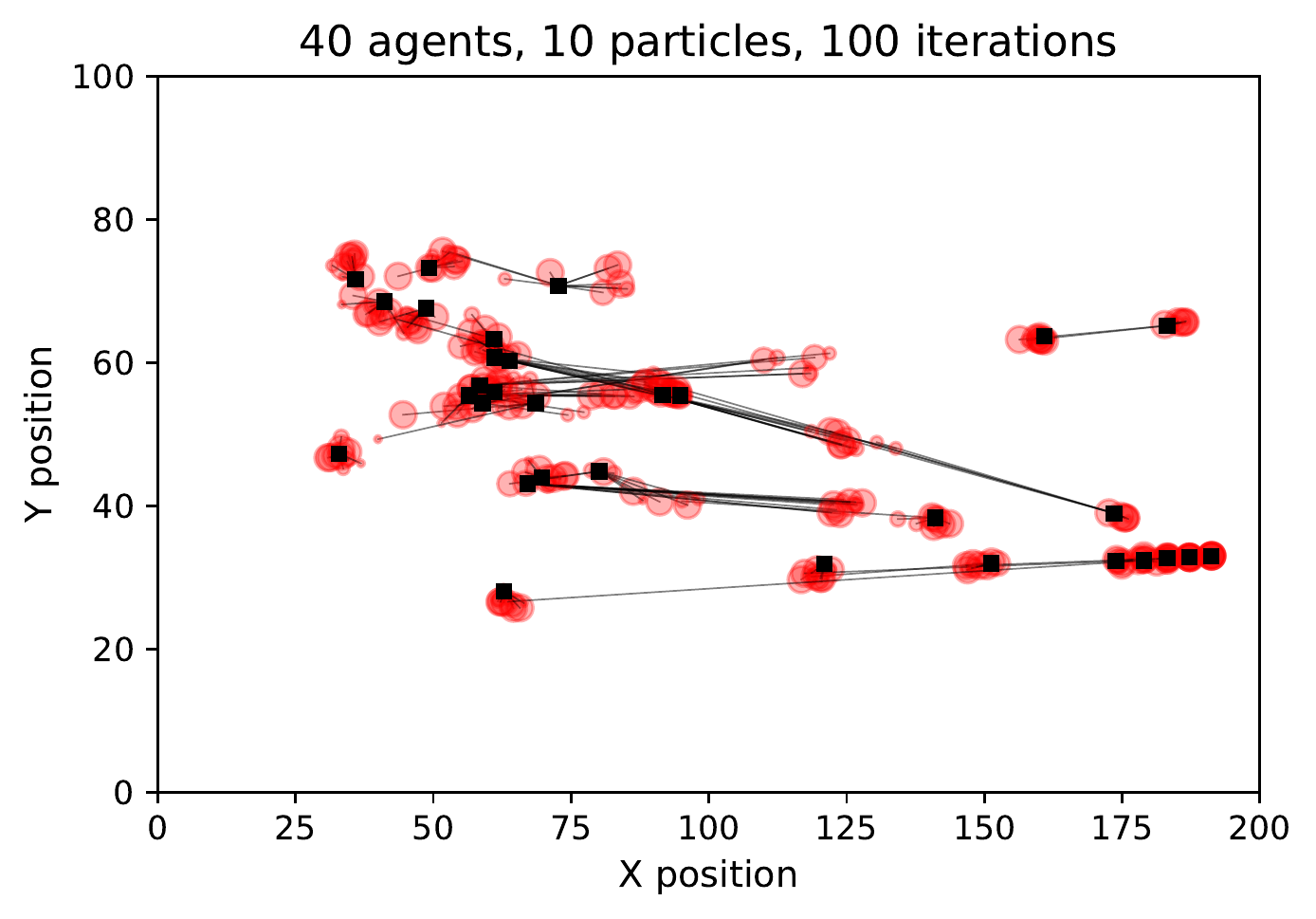}
	\includegraphics[width=0.45\textwidth]{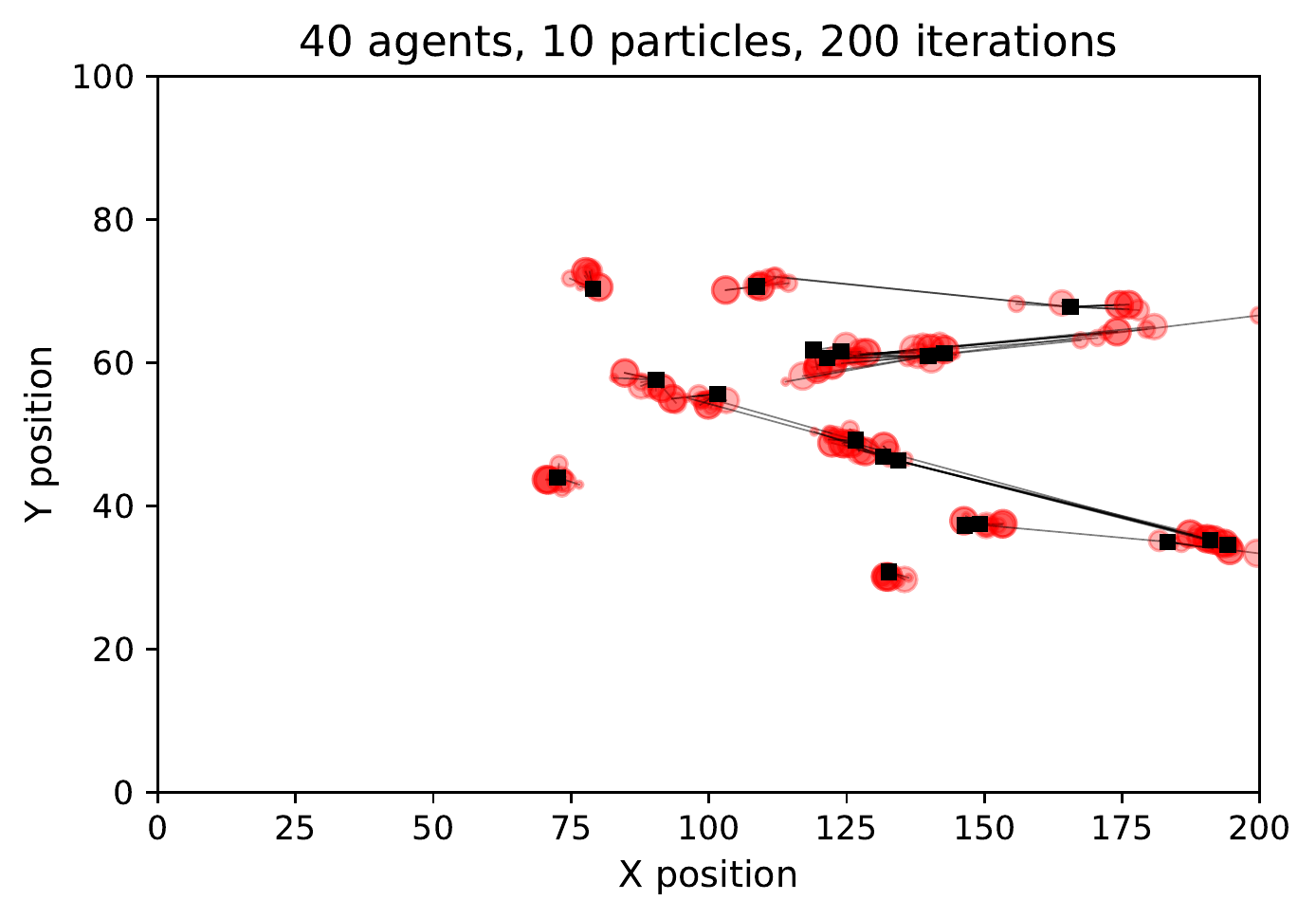}
	\caption{State of a particle filter with 50 agents and 10 particles. Lines connecting black squares (the pseudo-real agent locations) to red circles (particle estimate of the agent location) show that some particle estimates are a long way from the true agent locations.} \label{fig:ani-50agents-10particles}
\end{figure}

As the number of agents increases, collisions become much more likely and the chance of a particle diverging from the pseudo-truth state increases considerably. It becomes more common that no single particle will correctly capture the behaviours of \textit{all} agents. Therefore even after resampling there are some agents whose locations the particle filter is not able to accurately represent. Figure~\ref{fig:ani-50agents-10particles} illustrates a filter running with 40 agents and still only 10 particles. The long black lines show the locations of pseudo-real agents and the corresponding agents in the individual particles; it is clear that for some agents none of the particles have captured their pseudo-real locations. This problem can be mediated by increasing the number of particles. Figure~\ref{fig:median_abs_error} showed that with approximately 10,000 particles the error for simulations with 40 agents drops to levels that are comparable to the simulations of 10 agents. Hence a rule of thumb is that any particle filter with an overall error that is comparable to a clearly successful filter (i.e. Figure~\ref{fig:ani-10agents-10particles}) are reliably estimating the state of the system.

\subsubsection{Impact of Resampling}

\begin{figure}[ht]
	\centering
	\includegraphics[width=0.9\textwidth]{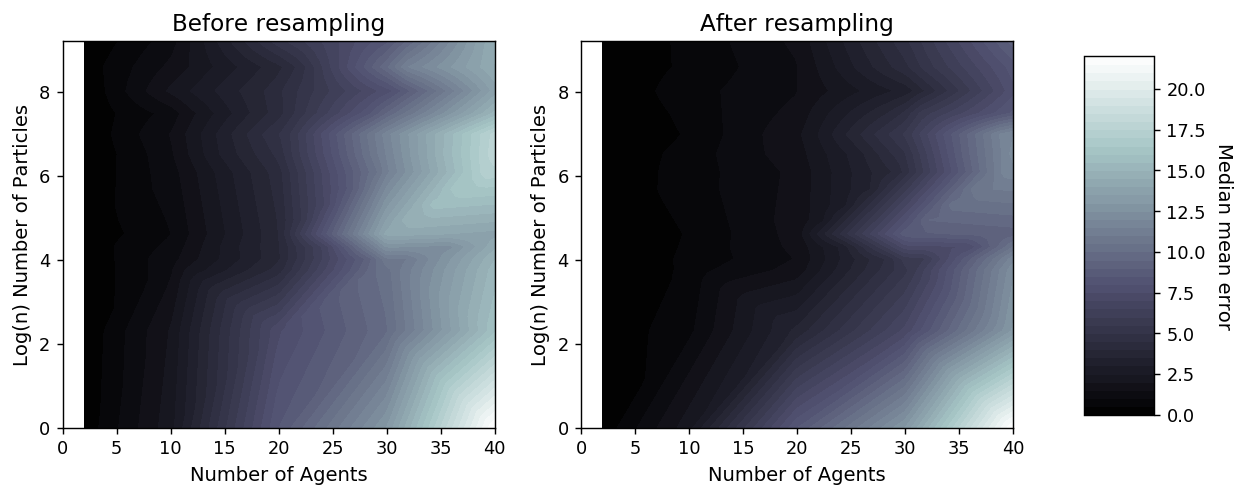}
	\caption{Median of the mean errors before and after resampling}\label{fig:resampling}
\end{figure}

Resampling is the process of weighting all particles according to how well they represent the pseudo-truth data; those with higher weights are more likely to be sampled and used in the following data assimilation window. This is important to analyse because it is the means by which the particle filter improves the overall quality of its estimates. Figure~\ref{fig:resampling} illustrates the impact of resampling on the error of the particle filter. With fewer than 10 agents in the system resampling is unnecessary because all particles are able to successfully model the state of the system. With more than 10 agents, however, it becomes clear that the population of particles will rapidly diverge from the pseudo-truth system state and resampling is essential for the filter manage to limit the overall error.

\subsubsection{Particle Variance}

As discussed in Section~\ref{particle_sampling}, the variance of the population of particles can be a reliable measure for estimating whether particle deprivation is occurring. If there is very little variance in the particles, then it is likely that they have converged close to a single point in the space of possible states. This needs to be avoided because, in such situations, it is extremely unlikely that any particles will reliably represent the state of the real system. Figure~\ref{fig:variance_results} illustrates this by visualising the mean error of all particles, $\nu(t)$ (defined in Equation~\ref{eqn:particle_error}), in each data assimilation window, $t$, and their variance under different numbers of agents and particles. Note that each agent/particle configuration is executed 10 times and the results are visualised as boxplots. Also, simulations with larger numbers of agents are likely to run for a larger number of iterations, but the long-running models usually have very few agents in them in later iterations (most have left, leaving only a few slow agents). Hence only errors up to 600 iterations, where most of the agents have left the environment in most of the simulations, are shown. The graphs in Figure~\ref{fig:variance_results} can be interpreted as follows: 

\begin{figure}[ht]
	\centering
	\includegraphics[width=0.9\textwidth]{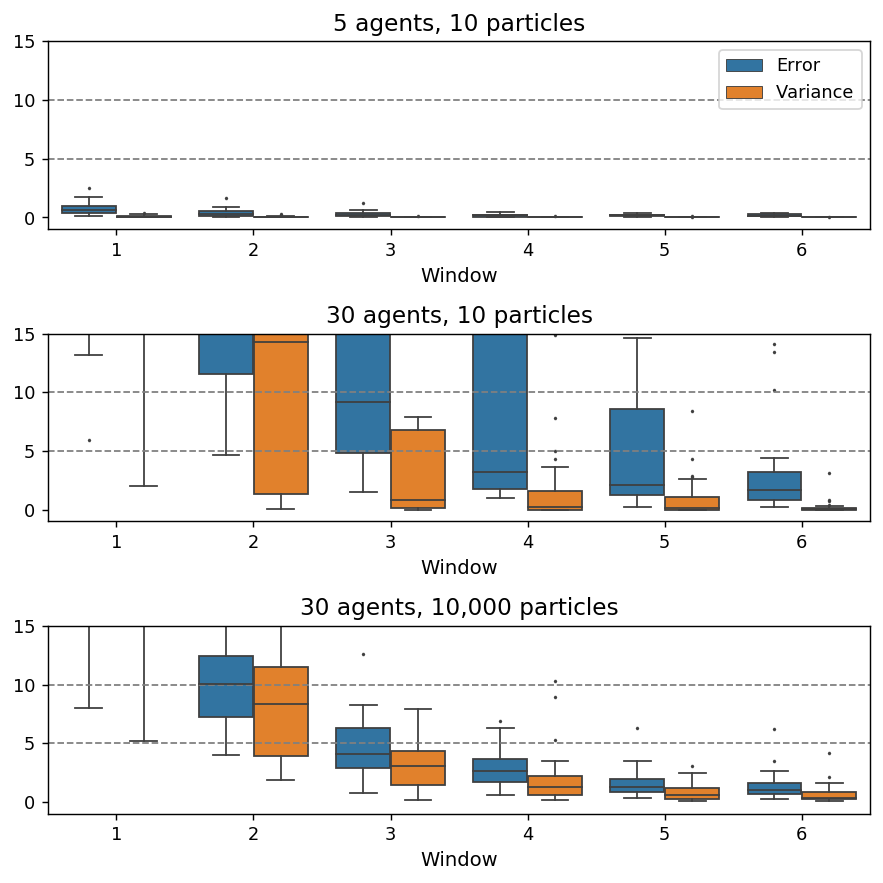}
	\caption{Mean error and variance of particles under different combinations of the number of particles and number of agents}\label{fig:variance_results}
\end{figure}

\begin{itemize}
    \item With \textbf{5 agents and 10 particles} there is very low error and low variance (so low that the box plots are difficult to make out on Figure~\ref{fig:variance_results}). This suggests that particle deprivation is occurring, but there are so few agents that the simulation is largely deterministic so the particles are likely to simulate the pseudo-truth observations accurately regardless.
    \item When the number of agents is increased to \textbf{30 agents and 10 particles}, the errors are much larger. The increased non-linearity introduced by the greater number of agents (and hence greater number of collisions) means that the population of particles, as a whole, is not able to simulate the pseudo-truth data. Although particle variance can be high, none of the particles are successfully simulating the target.
    \item Finally, with \textbf{30 agents and 10,000 particles}, the errors are relatively low in comparison, especially after the first few data assimilation windows. 
\end{itemize}

\subsubsection{Particle Noise}

\begin{figure}[ht]
	\centering
	\includegraphics[width=0.5\textwidth]{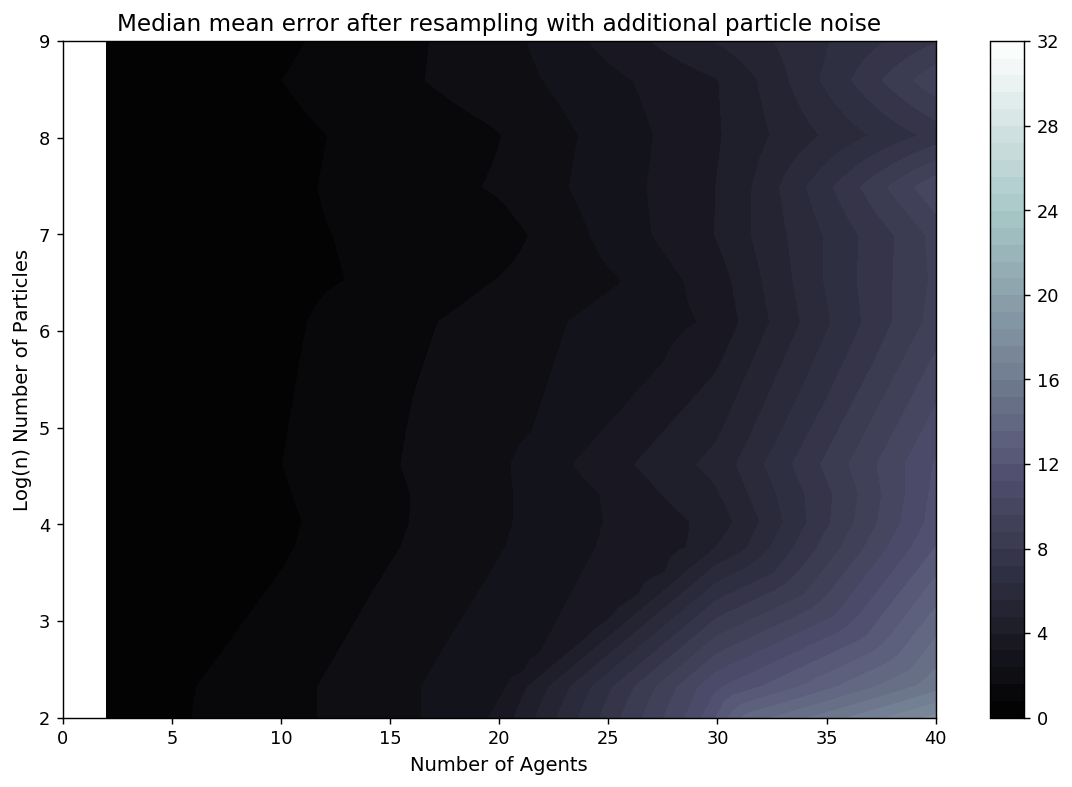}
	\caption{Median of the mean errors after resampling with additional noise ($\sigma=0.5$).} \label{fig:median_abs_error-noise2.0}
\end{figure}

The number of particles ($N_p$) is the most important hyper-parameter, but the amount of noise added to the particles ($\sigma_p$) is also important as this is the means by which particle deprivation is prevented. However, the addition of too much noise will push a particle a long way away from the true underlying state. Under these circumstances, although particle deprivation is unlikely none of the particles will be close to the true state. To illustrate this, Figure~\ref{fig:median_abs_error-noise2.0} presents the median error with a greater amount of noise ($\sigma_p=0.5$). The overall pattern is similar to the equivalent graph produced with $\sigma_p=0.25$ in Figure~\ref{fig:median_abs_error} -- as the number of agents in the simulation ($N_a$) increases so does the number of particles required to maintain low error ($N_p$)  -- but the overall error in each $N_a$/$N_p$ combination is substantially larger for the experiments with additional noise. Future work could explore the optimal amount of noise in more detail; $\sigma=0.25$ was found to be the most reliable through trial and error.

% !TEX root = ParticleFilter.tex
\section{Discussion and Future Work\label{discussion}}

\subsection{Discussion}

This paper has experimented with the use of a sequential importance resampling (SIR) particle filter (PF) as a means of dynamically incorporating data into a simple agent-based model of a pedestrian system. The results demonstrate that it is possible to use a particle filter to perform dynamic adjustment of the model. However, they also show that (as expected~\citep{rebeschini_can_2015, carrassi_data_2018}) as the dimensionality of the system increases, the number of particles required to maintain an acceptable approximation error grows exponentially. The reason for this is because, as the dimensionality increases, it becomes less likely that an individual particle will have the `correct combination' of values~\citep{long_spatial_2017}. In this work, the dimensionality is proportional to the number of agents. At most 10,000 particles were used, which was sufficient for a simulation with 30-40 agents. However, for a more complex and realistic model containing hundreds or thousands of agents, the number of particles required would most likely number in the millions. The particle filter used in this study was provided with more information than would normally be available. For example, information was supplied as fixed parameters on when each agent entered the simulation, their maximum speeds, and their chosen destinations. Therefore the only information that the particle filter was lacking was the actual locations of the agents and whether they would chose to move left or right to prevents a collision with another agent. It is entirely possible to include these agent-level parameters in the state vector, but this would further increase the size of the state space and hence the number of particles required. This is an important caveat as in a real-world situation it is very unlikely that such detail would be available. Future work should begin to experiment with the number of particles that would be required when observational data that are more akin to those available in the real world are used.

There are a number of possible improvements that could be made to the basic SIR particle filter to reduce the number of particles required. For example, \citep{wang_data_2015} propose \textit{component set resampling} -- details below -- but exploring these further is beyond the scope of this paper. Overall, these results reflect those of other studies which demonstrate that particle filtering has value for simulations with relatively few agents and interactions \citep[e.g.][]{wang_data_2015, lueck_who_2019, kieu_dealing_2019} but that the large dimensionality of a pedestrian system poses problems for the standard (unmodified) bootstrap filter. 

\subsection{Improvements to the particle filter}

There are a number of adaptions to the main particle filtering algorithm that might make the method more amenable to use with complex agent-based models. The aforementioned Component Set Resampling \citep{wang_data_2015} approach proposes that individual components of particles are sampled, rather than whole particles in their entirety. A more commonly used approach is to reduce the dimensionality of the problem in the first place. With spatial agent-based models, such as the one used here, spatial aggregation provides such an opportunity. In the data assimilation stage, the state vector could be converted to an aggregate form, such as a population density surface, and particle filtering could be conducted on the aggregate surface rather than on the individual locations of the agents. After assimilation, the particles could be disaggregated and then run as normal. This will, of course, introduce error because the exact positions of agents in the particles will not be known when disaggregating, but that additional uncertainty might be outweighed by the benefits of a more successful particle filtering overall. In addition, the observations that could be presented to an aggregate particle filter might be much more in line with those that are available in practice (as will be discussed shortly). If the aim of real-time model calibration is to give decision makers a \textit{general idea} about how the system is behaving in real time, then this additional uncertainty might not be problematic. A similar approach, proposed by \citep{rebeschini_can_2015}, could be to divide up the space into smaller regions and then apply the algorithm locally to these regions (i.e. a divide-and-conquer approach). Although it is not clear how well divide-and-conquer would work in an agent-based model -- for example, \citep{long_spatial_2017} developed the method for a discrete cellular automata model -- it would be an interesting avenue to explore.

\subsection{Real-World Implications and Future Work}

It is important to note that, unlike other data assimilation approaches, the particle filter does not dynamically alter the state of the running model. This could be advantageous because, with agent-based modelling, it is not clear that the state of an agent should be manipulated by an external process. Agents typically have goals and a history, and behavioural rules that rely on those features, so artificially altering an agent's internal state might disrupt their behaviour making it, at worst, nonsensical. Experiments with alternative (potentially more efficient) algorithms such as 4DVar or a the Ensemble Kalman Filter should be conducted to test this. 

Ultimately the aim of this work is to develop methods that will allow simulations of human pedestrian systems to be optimised in real time. Not only will such methods provide decision makers with more accurate information about the present, but they could also allow for better predictions of the near future \citep[c.f.][]{kieu_dealing_2019}. One assumption made throughout the paper, which limits its direct real-world use, is that the locations of pseudo-real individuals are known, albeit with some uncertainty. Not only is this assumption unrealistic -- it is rare for individuals to be tracked to this degree in public places -- but we would also argue that the privacy implications of tracking individual people are not outweighed by the benefits offered by better understanding the system. Therefore, immediate future work will test how well a data assimilation algorithm would fare were it supplied only aggregate information such as the number of people who pass through a barrier, or the number of people recorded by a CCTV camera within a particular area. Both of these measures can be used in aggregate form and would be entirely anonymous. It is unclear whether such aggregate data would be sufficient to identify a `correct' agent-based model, so experiments should explore the spatio-temporal resolution of the aggregate data are required. Also, identifiability/equifinality analysis might help initially as a means of estimating whether the available data are sufficient to identify a seemingly `correct' model in the first place. In the end, such research might help to provide evidence to policy makers for the number, and characteristics, of the sensors that would need to be installed to accurately simulate the target system, and how these could be used to maintain the privacy of the people who they are recording.

\bibliographystyle{unsrt}
\bibliography{2018-ParticleFilter}

\begin{appendices}
    \section{Instructions for Running the Source Code\label{appendix:code}}

The source code to run the StationSim model and the particle filter experiments can be found in the main `\href{https://dust.leeds.ac.uk/}{Data Assimilation for Agent-Based Modelling}' (DUST) project repository:
\begin{quote}
    \texttt{
        \href{https://github.com/urban-analytics/dust}
        {github.com/urban-analytics/dust}
    }
\end{quote}

Specifically, scripts and instructions to run the experiments are available at: 
\begin{quote}
    \texttt{
        \href{https://github.com/Urban-Analytics/dust/tree/master/Projects/ABM_DA/experiments/pf_experiments}
        {github.com/Urban-Analytics/dust/tree/master/Projects/ABM\_DA/experiments/pf\_experiments}
    }
\end{quote}

\end{appendices}

\end{document}